\documentclass{article}
\usepackage{graphicx}

\begin{document}
\title{ (3+0)D electromagnetic solitons and de Broglie's ''double solution''.
\author{Jacques Moret-Bailly 
\footnote{Laboratoire de physique, Université de Bourgogne, BP 47870, F-21078 Dijon cedex, France. email : 
Jacques.Moret-Bailly@u-bourgogne.fr}}}
\maketitle

\begin{abstract}
	The well known light filaments are obtained in various media whose index of refraction increases before a 
saturation with the electric field; adding a small perturbation which increases the index  with the magnetic field, and 
neglecting the absorption, a filament curves and closes into a torus. This transformation of a (2+1)D soliton into a 
(3+0)D soliton shows the existence of those solitons, while a complete study, with a larger magnetic effect, would 
require numerical computations, the starting point being, possibly, the perturbed, curved filament.

	The flux of energy in the regular filaments is nearly a ''critical flux'', depending slightly on the external fields, so 
that the energy of the (3+0)D soliton is quantified, but may be slightly changed by external interactions.

	The nearly linear part of the soliton, an evanescent wave, is partly transmitted by Young holes, making 
transmitted and reflected interference patterns, thus index variations which guide the remainder of the soliton, just as de 
Broglie's pilot waves.

	The creation of electron positron pairs in the vacuum by purely electromagnetic fields shows a nonlinearity of 
vacuum at high energies; supposing this nonlinearity convenient, elementary particles may be (3+0)D solitons or light 
bullets, so that it may be a connection with the superstrings theory.

Keywords : Wave-particle duality, Solitons.
Pacs :03.65.Bz, 03.50.De, 42.65.Tg 
\end{abstract}
\maketitle

\section{Introduction}
The old concept of classical point-particle leads to infinite electromagnetic and gravitational energies. Up to now, the 
most efficient models of particle are non-classical, widely phenomenological: Dirac's electron, su(n) algebra, skyrmions 
\cite{Skyrme,Gisiger}, solutions of Maxwell-Dirac equations \cite{Rosen, Lisi} for instance; these two last methods use 
solitons to avoid the infinite energies of waves near singular points. Born and Infeld \cite{Born} proposed to replace 
Maxwell's equations by a nonlinear system of equations; while optics in the vacuum is perfectly linear, this hypothesis 
is not absurd being a classical alternative to the virtual particles which allow the creation of electron-positron pairs in 
the vacuum \cite{Ritus}.

We have no powerful mathematical theory to study nonlinear systems. However, a mixture of theory, experiments and 
numerical computations led to reliable, important results in the field of the optical solitons. Some results explained in 
reviews \cite{Stegeman,Kivshar}, or in papers cited therein are condensed, in section 2, into a set of properties making a 
theorem which describes the properties and the stability of (2+1)D \footnote{In this notation, the first number gives the 
dimension of constraint of the wave, the second the dimension of propagation} solitons in the optics of Kerr, 
photorefractive and similar nonlinear media.

Using the known results about these (2+1)D solitons, section 3 demonstrate the existence of (3+0)D solitons in an 
hypothetical medium whose index of refraction is an increasing function of both the square of the electric field and the 
square of the magnetic field, up to saturations. Although such a soliton goes through a single Young hole, it is subject 
to interferences, so that a wave particle duality appears.

In section 4 the first step of an identification of a (3+0)D soliton with de Broglie's double solution is proposed. 

\section{Some properties of the filaments of light}
In media whose index of refraction increases quickly with the amplitude of an electric field, up to a saturation, a powerful 
laser beam splits into filaments.
These filaments of light which are (2+1)D solitons, have been extensively studied experimentally and theoretically \cite{ 
Chiao,Marburger,Stegeman2,Brodeur,Feit,Zharov}. The required nonlinearity is found in many media, in particular Kerr, 
photorefractive, or plasma; here, we will consider the propagation of a monochromatic wave in perfect, homogenous, 
isotropic, lossless media.

Set $Oz$ the axis of a single filament, in the free space; almost all its energy propagates in a cylinder of axis $Oz$ named 
the core; outside, the nonlinearity is negligible and the field is evanescent. The flux of energy in the filament has a well 
defined "critical" value. If the space is limited, or if the evanescent wave is perturbed by an other field, the flux of energy 
is slightly modified. The wave is plane and its period defines a wavelength $\Lambda$.

In a lossless medium, the filament may be very stable; perturbations which have the cylindrical symmetry leave the 
filament nearly unchanged, but an unsymmetrical perturbation curves the filament without destruction although its 
sections perpendicular to the direction of propagation become nearly elliptical \cite{Petter}; the complicated interaction 
of a filament with an other filament \cite{Shih} curves the axis of a filament without destruction up to a spiralling of both 
filaments \cite{Poladian,Shih2,Belic}.

The perturbations provided by a field coherent with the field in the filament do not destroy the filament although the 
addition of an external field to the inhomogeneous field inside the filament changes the index of refraction much more 
near the centre of the filament than in nearly linear regions: the stability allows a simple, global refraction of the filament.

When a set of filaments appears from a focussed laser beam, or from a flat soliton, a burgeoning filament absorbs energy 
from its neighbourhood; the length of a filament in an absorbing medium shows that it absorb energy to maintain its 
energy near the critical value. On the contrary, focussing a very homogeneous beam may produces a too powerful 
filament which looses quickly its extra energy: there is an equilibrium between the flux of energy in the filament and the 
energy outside, the critical value corresponding to a filament in a free space.

The variation of the index of refraction depends on the square of the instantaneous value of the  electric field, so that 
the coherence of the field in the filament with the perturbing field is important; we add the intensities of the incoherent 
fields, the amplitudes of the coherent fields; for instance, depending on their relative phase, two filaments may attract or 
repulse each other.

A consequence of this last behaviour is an interference of a light filament with itself, probably too weak for an 
observation : if a filament crosses a hole of a screen, it loses a part of its evanescent wave. \footnote{This part may be 
recovered from the zero point field often named ''stochastic'' although it is only stochastic far from matter.}. If there is a 
second hole in the screen, a small part of the evanescent field crosses this second hole; after, it perturbs the filament, 
the coherence making the interaction relatively large and directing the filament to bright fringes: it is a Young's 
experiment.
\section{Perturbation of a filament by a magnetic nonlinearity.}
Set $\mbox{\boldmath $E$}(x, y, z, t)$ the electric field, assumed polarised along $Ox$, of the filament propagating along 
the $Oz$ axis, and $E(x, y, z, t)$ the non-zero component of this field. 

\medskip
The electric field $\mbox{\boldmath $E$}(x, y, z, t)$ is an exact solution of Maxwell's equations in which the relative 
permittivity $\epsilon$ is a function of $|\mbox{\boldmath $E$}(x, y, z, t)|$, and the relative permeability $\mu$ is 1. 
Neglecting $|\mbox{\boldmath $\nabla . E$}(x, y, z, t)|$, Helmholtz propagation equation is obtained
\begin{equation}
\mbox{\boldmath $\Delta E$}(x, y, z, t)=\frac{\mu\epsilon(|E(x, y, z, t)|)}{c^2}{\partial^2 {\bf  E}(x, y, 
z, t)\over\partial t^2}.\label{propa}
\end{equation}
Set \cite{Chiao}
\begin{equation}
\mbox{\boldmath $E$}(x, y, z, t)= \mbox{\boldmath $E$}_t(x,y)\cos(\phi-\omega t) \label{propag}
\end{equation}
with $\phi=kz $.
$ \mbox{\boldmath $E$}_t(x,y)$ verifies the radial equation of the filament and may be written $ \mbox{\boldmath 
$E$}_t(r)$ while $\cos(kz-\omega t)$ is the propagation term; the fields remain unchanged by an increase of the 
coordinate $z$ by a period $\Lambda$. Assuming that the beam is cylindrical, and that there is no absorption, the time-
reversal invariance shows that $ \mbox{\boldmath $E$}_t(x,y)$ is real, so that the wavefronts are planes perpendicular 
to $Oz$.

\medskip

Assume a perturbation such that the index of refraction $n(x,y)$, with $n^2(x,y)=\epsilon(x,y)\mu(x,y)$ increases not 
only with the electric field, but with the modulus of $|\mbox{\boldmath $\nabla \wedge E$}(x, y, z, t)|$ , that is with the 
time derivative of the modulus of the magnetic field (or with the modulus of the amplitude of magnetic field for a fixed 
frequency). The variation of the permeability may be written , using cylindrical coordinates $(r,\theta,z)$ : 
\begin{equation}
\delta\mu_0=f_0(({\rm curl} \mbox{\boldmath $E$}(r,\theta,z))^2)=f_0\Bigl\{\Bigl (\frac{\partial  \mbox{\boldmath 
$E$}}{\partial z}\Bigr)^2+
\frac{1}{r^2}\Bigl(\frac{\partial(r E_\theta)}{\partial r}+\frac{\partial E_r}{\partial \theta}\Bigr)^2\Bigr\}\label{mu1}
\end{equation}
Respecting the symmetry, this perturbation does not curve, or modify much, the filament.

\medskip
Consider now an other problem, in the same homogenous, isotropic medium; to set it, we use curved cylindrical 
coordinates described in the figure, $R$ being a parameter; with these coordinates, the components of the curl are:
\begin{figure}
\begin{center}
\includegraphics[height=5 cm]{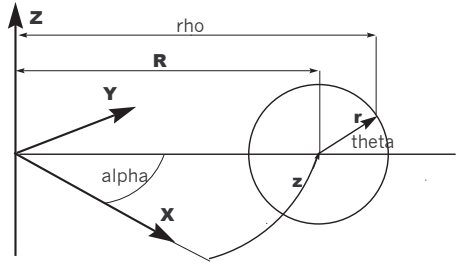}
\end{center}
\caption{Curved cylindrical coordinates}
\end{figure}
\begin{equation}
({\rm curl} \mbox{\boldmath 
$E$})_r=\frac{1}{r(R+r\cos\theta)}\Bigl\{\frac{\partial[(R+r\cos\theta)E_\alpha]}{\partial\theta}- \frac{r\partial 
E_\theta}{\partial \alpha}\Bigl\}
\end{equation}
\begin{equation}
({\rm curl} \mbox{\boldmath $E$})_\theta=\frac{1}{R+r\cos\theta}\Bigl\{\frac{\partial E_r}{\partial \alpha}-\frac{\partial 
[(R+r\cos\theta)E_\alpha]}{\partial r}\Bigr\}
\end{equation}
\begin{equation}
({\rm curl} \mbox{\boldmath $E$})_\alpha=\frac{1}{r}\Bigl\{\frac{\partial (r E_\theta)}{\partial r}-\frac{\partial 
E_r}{\partial \theta}\Bigr\}
\end{equation}
Suppose that a daemon sets electric and magnetic fields equals to the value they had in the previous problem, with the 
relation $z=R\alpha$. Taking into account that $E_\alpha$ is null, the perturbation induces a variation of the permeability
\begin{equation}
\delta\mu=f(({\rm curl} \mbox{\boldmath $E$}(r, \theta,\alpha))^2)=f\Bigl\{
\frac{1}{(R+r\cos\theta)^2}\Bigl (\frac{\partial \mbox{\boldmath $E$}}{\partial \alpha}\Bigr)^2+
\frac{1}{r^2}\Bigl(\frac{\partial(r E_\theta)}{\partial r}-\frac{\partial E_r}{\partial \theta}\Bigr)^2\Bigr\}\label{mu2}
\end{equation}

Equations \ref{mu1} and \ref{mu2} differ by a term, depending on $R/\rho$ with $\rho=R+r\cos\theta$, which produces 
a decrease of the index of refraction according to $\rho$. Thus, for a propagation of the wave surface corresponding to 
an increase $\Delta\alpha$ of $\alpha$, the wave surface is turned by an angle $\Delta\beta=\gamma\Delta\alpha$ which 
defines a  curvature $\gamma/R$. 

If $\gamma$ equals one, the daemon is useless, a toroidal solution is found. Is it stable? To ease the discussion, replace, 
in equation \ref{mu2}, $R$ by the inverse of the curvature $C=1/R$, so that $\delta\mu$ becomes a function of $r, \theta, 
\alpha$ depending on the parameter $C$. $\Delta\beta$ depends on the parameter $C$ and on $\Delta z= \Delta\alpha/C$.

Study the variation of $\Delta\beta(\Delta z, C)$ for a constant value of $\Delta z$. Choose function $f$ so that it 
increases fast for low values of $C$, then saturates; the variations of the index of refraction, and of $\Delta\beta$ are 
similar; thus, whichever its value for $C=0$, $\Delta\beta(\Delta z, C)$ becomes larger than $\alpha$ for a small value of 
$C$, then $\gamma$ decreases; thus, as $\Delta\beta$ reaches the value $\Delta\alpha$, the derivative ${\rm 
d}\Delta\beta/{\rm d}\Delta\alpha$ is lower than one; for the corresponding value $C_0=1/R_0$ of $C$ we have a locally 
stable solution.

The value $R_0$ of the radius of curvature depends only on the properties of the medium and of the wavelength, so that 
the kernel of the filament closes into a torus; the phases of the fields must be the same at the surface of junction; the 
phase may be adjusted, changing the frequency of the wave or the properties of the medium, so that $2\pi R_0$ 
becomes an integer multiple of $\Lambda$; thus, several toroidal solutions of the nonlinear Maxwell's equations are 
found. The demonstration requires that the evanescent field may be neglected for $\rho$ small. 

The flux of energy in the filament being near the critical value, and the length of filament transformed into a torus 
depending on the assumed properties of the medium, the energy of the soliton is quantified. The (2+1)D soliton is 
transformed into a (3+0)D soliton whose core occupies a limited region of the space, static, or, changing the galilean 
frame of reference, having any speed, on the contrary of light bullets ((3+1)D solitons) which move fast.

If the torus moves in relation to a screen and crosses the screen, its evanescent field is cut up and makes interferences; 
these interferences modify the index of refraction, so that the trajectory of the torus is perturbed.

\medskip

The existence of  electromagnetic (3+0)D solitons has been shown, apparently for the first time, in the particular case 
where the variation of the index of refraction produced by the magnetic field is low; it may be a starting point for 
numerical computations of more general solitons; numerical computations seem necessary to answer many questions 
such as:

- what happens increasing the magnetic nonlinear contribution to the index of refraction: in particular, can the torus 
become next to a sphere ?

- can the soliton have an electric or magnetic charge introduced by a non zero divergence of the fields ?

- how does the cutting of the evanescent wave precisely modifies the trajectory of the torus?

\medskip

The process of building (3+0)D solitons may be extended, for instance if the function which represents the variation of 
the index of refraction according to the fields has several maximums, or adding a torsion to the curvature.
\section{Wave particle duality of (3+0)D solitons}
These solitons do not seem useful in regular optics because magnetic crystals, such as tourmaline absorb the light 
much. The balls of fire produced by the lightnings may be solitons in ionised gases.

As the trajectory of a soliton is perturbed by a cutting of the evanescent field, this field is similar to de Broglie's pilot 
field $\Psi$ while the remainder plays the role of his $u$ field \cite{deBroglie}. This purely classical wave-particle duality 
is consistent with de Broglie's trials.

In the vacuum, up to X rays, Maxwell's equations with linear parameters are well verified, but a powerful enough 
$\gamma$ photon interacts with an electric or magnetic field, or with an other $\gamma$ photon to produce an electron 
pair \cite{Ritus}. In quantum physics, the required nonlinearity is introduced through virtual particles; in a classical 
scheme, nonlinear terms must be introduced in Maxwell's equations (Schwinger \cite{Schwinger}); this introduction 
breaks the superposition property of linear systems, gives individualities to regions of field such as (3+0)D solitons. The 
properties of the ring model of the electron \cite{Allen} bound to its symmetries could apply to the (3+0)D soliton.

\section{Conclusion}
This probably first (3+0)D soliton should be generalised in non-perturbative conditions, but this aim seems to require 
big numerical computations.

Assuming a magnetic and electric nonlinearity of vacuum at high energy is a simple hypothesis which could give 
interesting results: A connection with the superstrings theory seems possible, the electromagnetic structure replacing 
the local topological structure; the neutrinos would be light bullets, necessarily fast. Fred Hoyle's continuous creation 
of matter could be a transformation of high frequency zero point waves into solitons, solving simultaneously the UV 
divergence.

\end{document}